\documentclass[aps,pre,reprint,superscriptaddress,floatfix]{revtex4-2}

\usepackage[utf8]{inputenc}
\usepackage[T1]{fontenc}
\usepackage[english]{babel}
\usepackage{microtype}

\usepackage{amsmath,amssymb,bm}
\usepackage{graphicx}
\usepackage{booktabs}
\usepackage{placeins}

\usepackage{xcolor}
\definecolor{neonblue}{HTML}{006BFF}
\usepackage{hyperref}
\hypersetup{
    colorlinks=false,
    citebordercolor={0 1 0},
    linkbordercolor={1 0.2 0.2},
    urlbordercolor={0 0.42 1},
    pdfborder={0 0 1}
}

\newcommand{\ta}{\tau_{\alpha}}
\newcommand{\fR}{f_R}
\newcommand{\fP}{f_P}
\newcommand{\ph}{\phi}
\newcommand{\DTR}{D_{\mathrm{TR}}}
\newcommand{\DPR}{D_{\mathrm{PR}}}

\begin{document}

\title{Time reparametrization across kinetic rules in a glass-forming liquid}
\author{Leonardo dos Santos Vitoria}
\affiliation{Physics Department, Federal University of Lavras, Lavras, Minas Gerais, Brazil}

\begin{abstract}
Monte Carlo algorithms combining swap moves with kinetic constraints
allow the relaxation pathways of a liquid to be modified without
changing its equilibrium measure. Recent work suggests that much
of this change can be absorbed into a time reparametrization,
yielding a near-collapse of parametric relations between overlaps
measured at different spatial scales. We examine this correspondence
in a three-dimensional polydisperse hard-sphere liquid by comparing
a planar $\fP$ rule that confines the translation of a fraction of particles
to randomly oriented planes with a pure-$\fR$ rule that restricts
swap participation. At each packing fraction, the two rules are
tuned to comparable structural relaxation times, and time is
eliminated explicitly. For $0.58\leq\ph\leq0.62$, we find a
near-collapse accompanied by a small but resolved residual that
persists when directly confined particles are excluded from the
observable. A \textit{self}--\textit{distinct} decomposition shows
that the statistics of individual displacements reproduce most
of the scale of this residual. Among translationally free particles,
the probability of large displacements decreases as the relative
directions to constrained neighbors become more aligned with their
forbidden normals, even when the number of constrained neighbors
is held fixed. This association is consistent with zero in a
\emph{ghost} control, where the same normals impose no translational
restriction. Our findings complement the time-reparametrization
picture by showing that a near-collapse of collective parametric
relations can coexist with rule-dependent displacement statistics
and an orientational signature in the mobility of particles that
are not directly constrained.
\end{abstract}

\maketitle

\section{Introduction}

The dynamics of a supercooled liquid, or more generally a glass former, depends, on the one hand, on its equilibrium organization and, on the other, on the microscopic rules governing its evolution in time \cite{BerthierBiroli2011}. This distinction can be explored particularly directly in simulations in which the dynamics is modified without changing the stationary distribution of the system. Several important approaches exploit precisely this possibility, for instance, swap Monte Carlo algorithms \cite{GrigeraParisi2001,BerthierCoslovichNinarelloOzawa2016,
NinarelloBerthierCoslovich2017}; event-chain algorithms \cite{BernardKrauthWilson2009,MaggsKrauth2022,
GhimentiBerthierVanWijland2024}; and irreversible samplers based on transverse forces \cite{GhimentiEtAl2023,GhimentiEtAl2024Transverse}. These different choices for manipulating the dynamics can substantially alter how configuration space is explored while preserving the equilibrium measure of the liquid \cite{BerthierGhimentiVanWijland2024}. In this sense, choosing the dynamical rule makes it possible to change the kinetics in a controlled way and study which aspects of relaxation emerge from the configurational organization of the liquid and which still depend on how the system moves between configurations.

A simple possibility is that each kinetic rule provides a different
clock for the same slow dynamics
\cite{ChamonEtAl2002,Kurchan2023,Kurchan2024}. In this scenario,
changing the rule speeds up or slows down structural relaxation,
thereby changing $\ta$, without modifying the relations between
the functions that track this evolution
\cite{Kurchan2024,GhimentiEtAl2026}. The idea has its roots in the
time-reparametrization invariance developed for the slow dynamics
of glassy systems
\cite{CugliandoloKurchan1993,CugliandoloKurchan1994,
ChamonEtAl2002,ChamonEtAl2004,ChamonCugliandolo2007,Kurchan2023}
and is connected to later formulations based on effective clocks
and material time
\cite{DouglassDyre2022,RiechersEtAl2022,BohmerEtAl2024}.
If the hypothesis holds, two rules may produce very different
relaxation curves as functions of time while preserving the same
relation between the observables that track their evolution.
A direct test is to eliminate time itself and plot one relaxation
function against another. Ghimenti \emph{et al.} applied this
procedure to overlap functions measured at different scales.
Although the time-dependent curves change substantially across
different dynamics, the parametric plots exhibit an approximate
collapse once time is eliminated, a behavior the authors describe
in finite dimensions in terms of \emph{time-reparametrization
softness} \cite{GhimentiEtAl2026}. 

This correspondence, however, does not appear to be general, and
its limits have yet to be fully established. Even when the collapse
between observables is very good, one must still ask how far the
correspondence remains valid as the kinetic rule is changed. The
time-reparametrization test itself includes cases where it clearly
fails, such as kinetically constrained models and East-like dynamics
modified by swap moves
\cite{GutierrezGarrahanJack2019,GhimentiEtAl2026}. Studies of swap
algorithms further show that changing the global relaxation time
does not necessarily exhaust the changes produced in the microscopic
organization of mobility \cite{ShiraishiBerthier2024}. This motivates
a more specific question: can deviations from a common parametric
relation emerge within the same particulate liquid when distinct
kinetic rules act through different microscopic mechanisms?

The family introduced by Gavazzoni, Brito, and Wyart is particularly
well suited to this test \cite{GavazzoniBritoWyart2024}. In their
construction, the system can be kinetically perturbed in two ways: \textit{(i)} a fraction $\fR$ of the particles is prevented from participating in swap moves and \textit{(ii)} a fraction $\fP$ has its translation restricted by forbidding motion along a random direction associated with each particle, which in three dimensions corresponds to motion on randomly oriented planes. Despite strongly modifying the dynamics, these constraints preserve the same equilibrium measure
of the liquid while allowing the rate of relaxation to be varied
continuously \cite{GavazzoniBritoWyart2024}. The two interventions
therefore act on qualitatively distinct kinetic channels, one on
the degrees of freedom associated with swap and the other directly
on translation. This distinction is also related to the
interpretation that the acceleration produced by swap depends on
the additional channels of motion made available by the dynamical
rule \cite{WyartCates2017,BritoLernerWyart2018}. Within the same
liquid and at the same state point $\phi$, this family allows us to choose
rules with comparable relaxation times $\ta$ and examine whether their
parametric relations remain the same once their global relaxation
times have been matched. If two distinct rules are tuned to produce
the same structural relaxation time, $\ta^A \simeq \ta^B$, is this
sufficient for observables measured at different scales to preserve
the same relation to one another when explicit time is eliminated?

In this work, we investigate this question by comparing kinetic
rules with comparable structural relaxation times and eliminating
time through parametric relations between collective overlaps
measured at different spatial scales. A separation between these
relations reveals a change in the shape of relaxation that cannot
be absorbed into a change of a scalar clock. The planar constraint
also provides a microscopic probe of this difference. Each directly
constrained particle has a quenched direction along which it cannot
move, whereas the remaining particles are translationally free
in three dimensions. We therefore examine whether a residual
separation persists in this free population and whether its
mobility retains a dependence on the directional geometry of
the kinetic rule.

\section{Model and observables}

\subsection{Liquid and kinetic rules}

We consider a three-dimensional liquid of $N=512$ polydisperse hard spheres in a periodic cubic box. Diameters are drawn from a uniform distribution rescaled so that $\langle\sigma\rangle=1$, and the packing fraction $\ph$ controls the state point. The system size matches that used by Gavazzoni, Brito, and Wyart in three dimensions to explore extensively the family of kinetic rules, with small finite-size effects over the dynamical range considered in that study \cite{GavazzoniBritoWyart2024,BerthierEtAl2012}. The diameter distribution has a coefficient of variation of approximately $0.23$. Initial configurations at $\ph=0.58$ are prepared by low-density insertion, gradual isotropic compression, and relaxation; denser states are obtained by incremental compression of equilibrated configurations.

The dynamics combines translational displacements and diameter swaps. One sweep corresponds to $N$ elementary attempts; swap moves are proposed with probability $p_{\rm swap}=0.20$, following established protocols for polydisperse liquids \cite{NinarelloBerthierCoslovich2017,GavazzoniBritoWyart2024}. The family of rules is labeled by $\kappa=(\fR,\fP)$. Here, $\fR$ is the fraction of particles that cannot participate in swap moves, whereas $\fP$ is the fraction whose translational displacements are projected onto random quenched planes \cite{GavazzoniBritoWyart2024}. The constrained particle identities and plane normals remain fixed across the ten trajectories of a given configuration.

To make the comparison clearer, we distinguish three kinetic rules throughout. The \emph{reference}, $\kappa_0=(0,0)$, imposes no planar constraint and keeps all particles available for swap; it serves as a common reference for the parametric comparison. The \emph{planar} rule, $\kappa_P=(0,0.5)$, retains swap moves but translationally confines half of the particles to fixed random planes. The \emph{pure-$\fR$} rule, $\kappa_R=(\fR^*,0)$, prevents a fraction $\fR^*$ of the particles from participating in swap moves. The final production runs cover $\ph=0.58$, $0.59$, $0.60$, $0.61$, and $0.62$. This range allows us to follow what happens as the dynamics becomes slower and check whether the results remain consistent as $\ph$ is varied. At each state point, $\fR^*$ is tuned to match the structural relaxation time of the pure-$\fR$ rule to that of the planar rule. The values of $\fR^*$ are $0.660$, $0.635$, $0.605$, $0.580$, and $0.635$ for $\ph=0.58$, $0.59$, $0.60$, $0.61$, and $0.62$, respectively. The resulting relaxation times differ by no more than $5.6\%$ at any state point. The reference rule, $\kappa_0$, is not matched in $\ta$ to either
of these dynamics; it serves only as the common reference used to quantify their departures through $\DTR$.

For each state point and rule, we use $20$ initial configurations and generate $10$ independent dynamical trajectories from each. The quantities of interest are first averaged over the $10$ trajectories sharing the same initial configuration and only then compared across the $20$ configurations. In this way, we  can distinguish what is reproducible from the initial configuration from fluctuations specific to individual trajectories in an isoconfigurational viewpoint \cite{WidmerCooperHarrowellFynewever2004,BerthierJack2007}.

We evaluated $\ta$ using the neighborhood-persistence correlation,
\begin{equation}
 C(t)=\frac{1}{N}\sum_{i=1}^{N}\frac{1}{n_i}
 \sum_{j\in\mathcal N_i(0)}
 \Theta\!\left(0.50-\left|\Delta\bm r_j(t)-\Delta\bm r_i(t)\right|\right),
 \label{eq:neighborhood_C}
\end{equation}
where $\Theta$ denotes the Heaviside step function, $\Delta\bm r_i(t)=\bm r_i(t)-\bm r_i(0)$ is the displacement of particle $i$, and $\mathcal N_i(0)$ contains its $n_i$ initial neighbors within distance $1.40$. We define $\ta$ by $C(\ta)=e^{-1}$ \cite{EavesReichman2009,KarmakarParisi2013} and locate the crossing by linear interpolation in $\log t$. This neighborhood-persistence observable is related to cage- and neighbor-relative measures \cite{RabaniGezelterBerne1997,VivekKelleherChaikinWeeks2017,ShibaKeimKawasaki2018}, while local rearrangements provide complementary probes \cite{CandelierEtAl2010}.

\subsection{Multiscale overlap and removal of the clock}

We use the collective real-space overlap to compare relaxation at different spatial scales \cite{GuiselinTarjusBerthier2020,GhimentiEtAl2026}. For a spatial tolerance $a$, it is defined as
\begin{equation}
 F_o(a,t)=\frac{1}{N}\sum_{ij}
 \Theta\!\left(a-\left|\bm r_i(t)-\bm r_j(0)
 -\Delta\bm r_{\rm cm}(t)\right|_{\rm PBC}\right),
 \label{eq:Fo}
\end{equation}
where $\bm r_i(t)$ is the position of particle $i$ at time $t$, $\Delta\bm r_{\rm cm}(t)=N^{-1}\sum_i\Delta\bm r_i(t)$ is the mean particle displacement, and the subscript ${\rm PBC}$ denotes the minimum-image distance under periodic boundary conditions. The sum measures the reoccupation of initially occupied regions, irrespective of particle identity. Even decorrelated configurations have a finite overlap due to random occupation \cite{GuiselinTarjusBerthier2020,GhimentiEtAl2026}. In three dimensions, this background is $q_r(a)=4\pi\rho a^3/3$, where $\rho=N/L^3$ is the number
density and $L$ is the box length. We therefore use the connected overlap
\begin{equation}
 Q_o(a,t)=\frac{F_o(a,t)-q_r(a)}{1-q_r(a)}.
 \label{eq:Qo}
\end{equation}
We normalize it as $\widehat Q_o(a,t)=Q_o(a,t)/Q_o(a,0)$, with $Q_o(a,0)$ equal to unity to numerical precision in the final data. We consider $a=0.15$, $0.20$, and $0.30$.

For each pair $a_2>a_1$, we eliminate time by plotting one overlap against the other,
\begin{equation}
 \widehat Q_o(a_2,t)=\mathcal F_\kappa^{(a_2|a_1)}
 \!\left[\widehat Q_o(a_1,t)\right].
 \label{eq:parametric}
\end{equation}
Here, $\mathcal F_\kappa^{(a_2|a_1)}$ is the parametric relation for rule $\kappa$ and the specified pair of probing lengths; we write $\mathcal F_\kappa$ when the pair is fixed. Rules related only by a monotonic change of clock have the same  parametric relation \cite{GhimentiEtAl2026}. The use of one relaxation function as an internal coordinate is also related to formulations based on effective clocks and material time \cite{HecksherOlsenDyre2015,MehriIngebrigtsenDyre2021,
PeredoOrtizEtAl2022,RiechersEtAl2022,GhimentiEtAl2026}.

We quantify departures from this correspondence using the reference rule $\kappa_0$. For each pair $(a_2,a_1)$, we define the mean absolute separation over the common parametric window
$0.25\leq x\leq0.75$ as
\begin{equation}
 \DTR(\kappa)=\frac{1}{\Delta x}\int_{0.25}^{0.75}
 \left|\mathcal F_\kappa(x)-\mathcal F_{\kappa_0}(x)\right|\,dx,
 \label{eq:DTR}
\end{equation}
where $x=\widehat Q_o(a_1,t)$ and $\Delta x=0.50$. This window excludes the initial high-correlation regime and the low-correlation tail, where finite-sampling noise becomes more relevant \cite{ZwanzigAilawadi1969, CharbonneauDyerLeeYaida2016}. Functional distances have also been used to quantify collapse in scaling analyses \cite{BhattacharjeeSeno2001}; Eq.~(\ref{eq:DTR}) defines the
measure used here. We evaluate the integral on a common grid and estimate its resolution floor by splitting the realizations into two halves \cite{LageCastellanosEtAl2019}.

To distinguish the direct contribution of planar confinement, we repeat the overlap analysis using only translationally free particles in the planar rule and random subsets of equal size in the rules with $\fP=0$. For each subset, both indices in Eq.~(\ref{eq:Fo}) are restricted to its particles, with the normalization and background adjusted to the subset size. The confined particles remain present in the dynamics and are excluded only from the observable.

\subsection{Direct separation and van Hove decomposition}

Because $\DTR$ measures each rule relative to $\kappa_0$, we must still check whether the difference between planar and pure-$\fR$ could depend on this choice of reference. To this end, we also directly measure the area between their parametric curves over the same window as before,
\begin{equation}
 \DPR=\frac{1}{\Delta x}\int_{0.25}^{0.75}
 \left|\mathcal F_P(x)-\mathcal F_R(x)\right|\,dx.
 \label{eq:DPR}
\end{equation}
Thus, $\DPR$ directly quantifies the separation between the two rules matched in $\ta$, without using the reference dynamics in its definition.

Having established this direct comparison, we ask in which component of the motion the separation appears. The double sum defining $F_o$ can itself be decomposed into \textit{self} and \textit{distinct} terms, as in the standard decomposition of the van Hove function \cite{VanHove1954,KobAndersen1995}. The terms with $i=j$ track whether a particle remains within a distance $a$ of its own initial position, whereas the terms with $i\neq j$ describe the reoccupation of initial positions by other particles. We can therefore write exactly
\begin{equation}
 F_o(a,t)=F_s(a,t)+F_d(a,t),
 \label{eq:Fsum}
\end{equation}
where $F_s$ collects the terms with $i=j$ and $F_d$ those with $i\neq j$.

In isotropic three-dimensional systems, the \textit{self} component is directly related to the van Hove function \cite{VanHove1954},
\begin{equation}
 F_s(a,t)=4\pi\int_0^a r^2G_s(r,t)\,dr,
 \label{eq:vanhove_cumulative}
\end{equation}
In this decomposition, the displacement entering $G_s$ is evaluated
after subtracting the same mean particle displacement and using the
same minimum-image convention as in the collective overlap, so that
$F_o=F_s+F_d$ remains exact.
Thus, $F_s(a,t)$ is the cumulative probability that a particle undergoes a displacement smaller than $a$. Consequently, $F_s(a_2,t)-F_s(a_1,t)$ directly measures the fraction of particles with $a_1\leq\Delta r<a_2$. The distribution $G_s(r,t)$, and in particular its tails, is widely used to characterize heterogeneous motion in supercooled liquids \cite{KobDonatiEtAl1997,YamamotoOnuki1998,
StarioloFabricius2006,ScallietGuiselinBerthier2022}.

Keeping the normalization used for the collective overlap, we define
\[
Q_s(a,t)=\frac{F_s(a,t)}{1-q_r(a)}
\qquad\text{and}\qquad
Q_d(a,t)=\frac{F_d(a,t)-q_r(a)}{1-q_r(a)},
\]
so that
\[
Q_o(a,t)=Q_s(a,t)+Q_d(a,t)
\]
exactly. The self and distinct distances are evaluated at the same collective-overlap coordinate $x=\widehat Q_o(a_1)$ and serve as non-additive diagnostics of the separation.

\subsection{Displacements of translationally free particles at internal stages}

The self decomposition identifies which microscopic channel carries most of the separation, but does not directly show how the distribution of displacements is reorganized. To examine this question, we analyze only translationally free particles. For this population, the self overlap is
\begin{equation}
F_s^{\rm free}(a,t)=
\frac{1}{M}\sum_{i\in\mathcal S_{\rm free}}
\Theta\!\left(
a-\left|
\Delta\bm r_i^{\rm uw}(t)
-\Delta\bm r_{\rm cm}(t)
\right|
\right),
\end{equation}
where $\Delta\bm r_i^{\rm uw}(t)
=\bm r_i^{\rm uw}(t)-\bm r_i^{\rm uw}(0)$ is the unwrapped displacement of particle $i$, and
$\Delta\bm r_{\rm cm}(t)=N^{-1}\sum_{i=1}^{N}\Delta\bm r_i^{\rm uw}(t)$
is the mean displacement evaluated over all particles.
The set $\mathcal S_{\rm free}$ contains the $M$ particles without planar translational constraints; the same particle identities are used in the pure-$\fR$ comparison. We compare displacement distributions at internal stages defined by
\begin{equation}
F_s^{\rm free}(0.15,t_\kappa)=x,
\qquad x\in\{0.70,0.50,0.30\},
\label{eq:internalstage}
\end{equation}
where $t_\kappa=t_\kappa(x)$ depends on the rule. The comparison
therefore uses the same self-relaxation stage of the free population.
Throughout the free-particle analyses below, $\Delta r_i$ denotes
the magnitude
\[
\Delta r_i=
\left|
\Delta\bm r_i^{\rm uw}(t)-\Delta\bm r_{\rm cm}(t)
\right|.
\]
When the particle index is immaterial, we simply write $\Delta r$.

At each stage, we calculate the radial probability density of displacement magnitudes, $p^{\rm free}(r)=4\pi r^2G_s^{\rm free}(r)$, and the probability that a displacement exceeds the largest scale used in the parametric test,
\begin{equation}
P_\kappa^{\rm free}(x)=
P_\kappa\!\left(
\Delta r\geq0.30
\mid F_s^{\rm free}(0.15)=x
\right).
 \label{eq:tailprob}
\end{equation}
The difference
\begin{equation}
 \Delta P_{\rm tail}(x)=P_P^{\rm free}(x)-P_R^{\rm free}(x)
 \label{eq:dptail}
\end{equation}
measures how the planar rule changes the population of large displacements among particles with no translational constraint of their own. This interpretation follows the established use of the self van Hove function to reveal differences in the microscopic shape of relaxation, including between dynamics with comparable global relaxation scales \cite{ShiraishiBerthier2024,KobAndersen1995}.

\subsection{Orientational exposure and the \emph{ghost} control}

The planar constraint contains information absent from the pure-$\fR$ rule. Each constrained particle $k$ has a quenched normal $\widehat{\bm n}_k$ along which its translation is forbidden. To investigate whether this information remains visible in the dynamics of a free particle $j$, we consider its 12 nearest neighbors and define $n_C(j)$ as the number of these belonging to the planar-constrained population. For each constrained neighbor, let
\begin{equation}
 \widehat{\bm u}_{jk}=\frac{\bm r_k(0)-\bm r_j(0)}{|\bm r_k(0)-\bm r_j(0)|}
 \label{eq:ujk}
\end{equation}
denote the initial relative direction, evaluated under periodic boundary conditions using the minimum-image convention \cite{AllenTildesley2017, ViscardyGaspard2003}. The main orientational estimator is
\begin{equation}
 a_j=\frac{1}{n_C(j)}\sum_{k\in\mathcal C_j}\left|\widehat{\bm u}_{jk}\cdot\widehat{\bm n}_k\right|,
 \label{eq:aj}
\end{equation}
for $n_C(j)>0$, where $\mathcal C_j$ contains the constrained neighbors of $j$. Larger values of $a_j$ correspond to a geometry in which the separation $j$--$k$ has a larger component along the directions these neighbors cannot traverse.

Mobility is characterized by the event $m_j=\Theta(\Delta r_j-0.30)$. To separate orientation from the number of constrained neighbors, we estimate the dependence
\begin{equation}
 P(m_j=1\mid a_j,n_C)=b_{n_C}+\beta a_j,
 \label{eq:beta}
\end{equation}
allowing an intercept that depends on $n_C$. Thus, $\beta$ measures how the probability of a large displacement responds to local orientation after controlling for the number of constrained neighbors. This is a population-level association, rather than a deterministic propensity of an individual particle; the distinction matters in systems where dynamical fluctuations remain large even with the initial configuration held fixed \cite{BerthierJack2007}.

We also construct a \emph{ghost} control in the pure-$\fR$ dynamics, in which the same particle identities and quenched normals used in the planar geometry are retained solely as labels in the analysis, without imposing any additional restriction on displacements. This construction is conceptually similar to the use of \emph{ghost} particles by Krakoviack \cite{Krakoviack2014}, which preserve reference positions as an auxiliary device for measuring correlations in partially immobilized fluids. In the present case, the normals are used to calculate the same orientational descriptor $a_j$ in the pure-$\fR$ dynamics, although they have no effect on its motion. We denote the orientational slopes in the planar dynamics and the pure-$\fR$ \emph{ghost} control by $\beta_P$ and $\beta_R$, respectively, both estimated while controlling for the number $n_C$ of constrained neighbors. The coefficient $\beta_R$ thus measures the association between orientation and mobility that could arise from nominal geometry in the absence of the planar constraint. We quantify the contrast between the two rules as
\begin{equation}
\Delta\beta=\beta_P-\beta_R.
\label{eq:dbeta}
\end{equation}
We test the orientational association by randomizing the normals and permuting $a_j$ within groups with the same $n_C$.

The $95\%$ confidence intervals are obtained by bootstrapping configurations, preserving pairing by index when two rules are compared within the same state point. The initial configuration remains the independent statistical unit; the ten realizations are aggregated before resampling.

This construction separates the hypothesis into five successive questions. First, can we choose two physically distinct rules with $\ta^P\simeq\ta^R$? Second, once this global clock has been controlled, do the two rules also satisfy $\mathcal F_P(x)\simeq\mathcal F_R(x)$ when time is eliminated? Third, if a separation remains, does it belong to the pair of mechanisms themselves, and which part of the van Hove function carries it? Fourth, does this difference remain visible in the displacements of translationally free particles? Finally, does the vectorial geometry of the planar constraint remain imprinted on the mobility of these particles?

\section{Results and discussion}

The results follow the questions formulated above. We begin by checking the matching of the global clock and then explicitly eliminate time from the comparison. Starting from the difference that remains, we ask how much can be attributed directly to the confined particles, in which component of motion it is concentrated, and finally whether the directional geometry of the constraint can still be identified in the mobility of particles that remain translationally free.

\subsection{Comparable clocks do not imply the same multiscale trajectory}

Matching $\ta$ controls the global scale of relaxation but does not, by itself, establish dynamical equivalence between the two rules. At each packing fraction, the planar and pure-$\fR$ rules were chosen to have comparable structural relaxation times. The parametric test imposes a more stringent condition, because it checks whether this equivalence of global scales also extends to the relation between different displacement scales. If the difference between planar and pure-$\fR$ can be absorbed into a change of clock, then, upon reaching the same value $\widehat Q_o(a_1)=x$, the two rules must also have the same $\widehat Q_o(a_2)$ \cite{GhimentiEtAl2026}. Put simply, an exact reparametrization would make these curves coincide. In finite dimensions, however, the observed collapse is approximate \cite{GhimentiEtAl2026}, which leads naturally to the question we address here. If the curves do not coincide exactly, how large is the difference, and what physical information does it retain?

Displaying every combination of probing lengths and packing fractions would obscure the near-collapse and its small residual. For clarity, panels (a) and (b) show the widest probing-length pair, $(a_2,a_1)=(0.30,0.15)$, at the lowest and highest packing fractions. This pair yields the largest $\Delta\DTR$ among the three pairs examined at every state point; the intermediate packing fractions and alternative pairs remain included in the quantitative analysis.

\begin{figure*}[tb]
\centering
\includegraphics[width=0.94\textwidth]{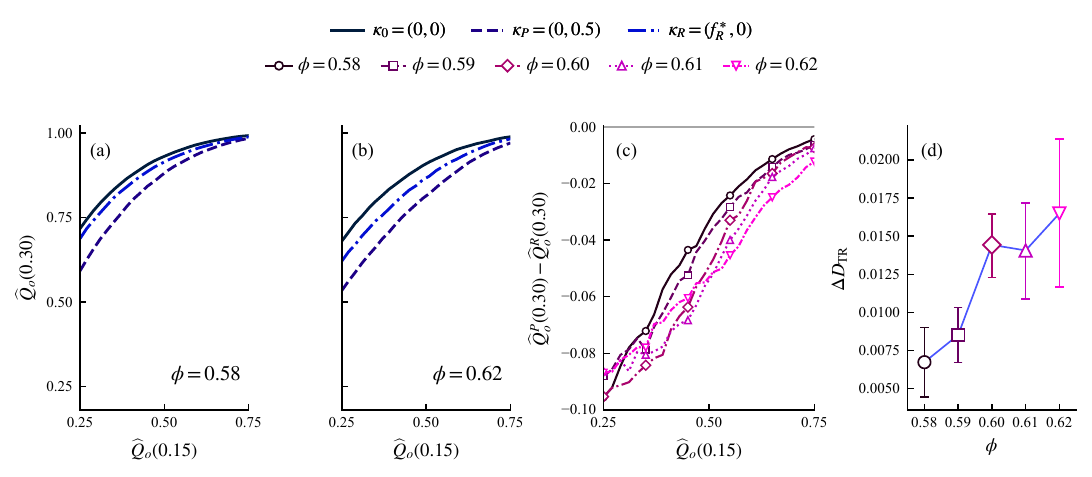}
\caption{\textbf{Time reparametrization and residual at the five state points.} (a,b) Parametric relations $\widehat Q_o(0.30)$ versus $\widehat Q_o(0.15)$ at $\ph=0.58$ and $0.62$, respectively, comparing the reference, planar, and pure-$\fR$ rules. (c) Direct difference $\Delta Q_{PR}(x)=\mathcal F_P(x)-\mathcal F_R(x)$ for all five packing fractions, with $x=\widehat Q_o(0.15)$. (d) Excess $\Delta\DTR=\DTR^P-\DTR^R$ for $(a_2,a_1)=(0.30,0.15)$ after the geometric control, with error bars indicating 95\% confidence intervals from resampling initial configurations.}
\label{fig:parametric}
\end{figure*}

Fig.~\ref{fig:parametric} shows that eliminating time brings the parametric relations produced by the different kinetic rules into close agreement. This \emph{near-collapse} is consistent with the \emph{time-reparametrization softness} observed in finite dimensions \cite{GhimentiEtAl2026}. However, the curves do not become exactly coincident. For the main pair $(a_2,a_1)=(0.30,0.15)$, planar and pure-$\fR$ remain very close throughout the analyzed window, yet retain a small systematic separation. This difference is seen more directly in $\Delta Q_{PR}(x)$, shown for all five state points in panel (c). Thus, matching $\ta$ and obtaining a good parametric collapse coexist with a residual difference between the two rules,
\begin{equation}
\ta^P\simeq\ta^R,\qquad
\mathcal F_P(x)\simeq\mathcal F_R(x),
\qquad
\mathcal F_P(x)-\mathcal F_R(x)\neq 0.
\label{eq:coexistence}
\end{equation}
In other words, controlling the global clock brings the two dynamics into close agreement, but information associated with the kinetic rule remains in the relation between different relaxation scales.

This separation can be tracked in an integrated manner through $\Delta\DTR=\DTR^P-\DTR^R$, shown in panel (d) of Fig.~\ref{fig:parametric}. Because $\DTR$ measures the departure of each parametric relation from the same reference, their difference allows us to compare how planar and pure-$\fR$ differ once time is no longer used as a coordinate. Over the range studied, $\Delta\DTR$ increases overall with $\ph$, although the values at $\ph=0.60$ and $0.61$ remain comparable. Thus, the data indicate that the residual does not decrease as relaxation becomes slower and suggest an increase in its magnitude between the less dense and more dense states investigated. However, the present set of state points does not allow us to establish a functional dependence on $\ph$. We therefore treat this evolution only as a trend observed over the accessible range, without assigning it an asymptotic form or a scaling law. At $\ph=0.61$, the alternative pairs $(0.20,0.15)$ and $(0.30,0.20)$ also yield positive $\Delta\DTR$, $0.0087$ and $0.0120$, respectively, with 95\% confidence intervals $[0.0070,0.0103]$ and $[0.0092,0.0149]$.

\subsection{The residual belongs to the mechanism and survives the geometric control}

\begin{figure*}[tb]
 \centering
 \includegraphics[width=0.94\textwidth]{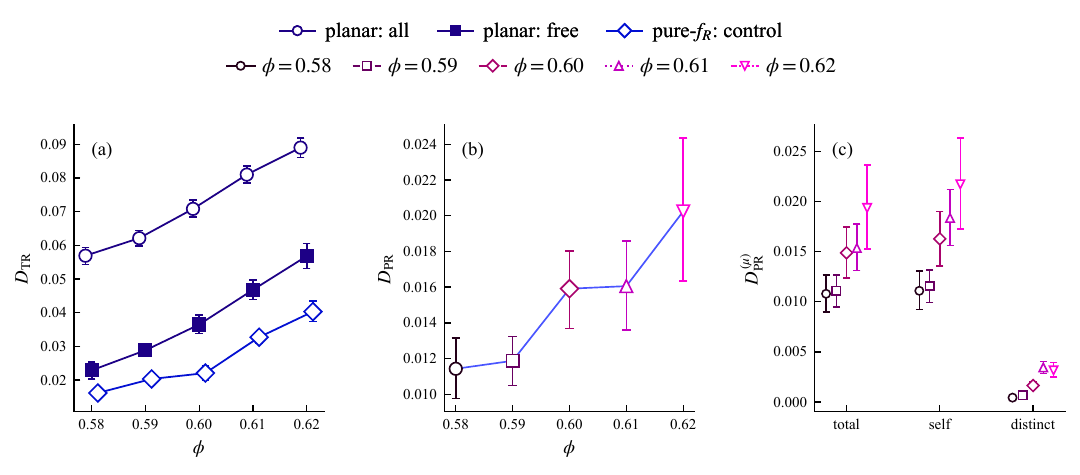}
 \caption{\textbf{Dissecting the residual after removal of the clock.}
 (a) $\DTR$ for the planar rule using all particles, the same
 observable restricted to translationally free particles, and
 the pure-$\fR$ control with matched subsets.
 (b) Direct distance $\DPR$ between planar and pure-$\fR$
 under the geometric control.
 (c) Total, self, and distinct diagnostics $D_{\rm PR}^{(\mu)}$
 at the five state points, where
 $\mu\in\{\mathrm{total},\mathrm{self},\mathrm{distinct}\}$.}
 \label{fig:origin}
\end{figure*}

The parametric curves in Fig.~\ref{fig:parametric}(a)--(c) are
constructed over the entire population. However, the planar rule
itself introduces a direct kinematic contribution that must be
separated before interpreting the origin of this residual. Half of
the particles explicitly lose one translational degree of freedom;
their contribution to $Q_o(a,t)$ can therefore produce part of the
observed difference even without any additional change in the
dynamics of particles that remain free. To check this possibility,
we repeat the calculation using only translationally free particles
and, for the rules with $\fP=0$, subsets of equal cardinality. As
shown in panel (a) of Fig.~\ref{fig:origin}, restricting the
observable in this way substantially reduces $\DTR$, confirming the
direct contribution of the confined population. Nevertheless, a
separation remains. In other words, the planar constraint also
modifies the relation between scales observed for particles that
remain free to translate in three dimensions.

We must then check whether this difference depends on the reference dynamics used in the definition of $\DTR$. To this end, we compare planar and pure-$\fR$ directly through $\DPR$, without using $\kappa_0$. Panel (b) of Fig.~\ref{fig:origin} shows that the separation remains resolved throughout the range studied and displays the same qualitative organization found in $\Delta\DTR$. Thus, the residual observed after removal of the clock does not arise solely from the choice of a common reference. It remains when the two matched rules are compared directly and, in this sense, is associated with the difference between the kinetic mechanisms themselves.

Having established that this separation persists both after the geometric control and in the direct comparison between the rules, we can ask in which component of motion it is concentrated. Using the exact decomposition $F_o=F_s+F_d$, the diagnostics shown in panel (c) indicate that the \textit{self} contribution has a magnitude comparable to the total separation, whereas the \textit{distinct} component remains considerably smaller. Reoccupation of initial positions by other particles therefore makes a measurable contribution, but most of the difference follows the statistics of the particles' own displacements. This shifts the interpretation of the residual toward the distribution of individual displacements and provides a starting point for examining directly how that distribution is reorganized between planar and pure-$\fR$.

\subsection{The self response persists among translationally free particles}

The preceding decomposition identifies which channel carries most of the separation, but does not yet show how the displacements are reorganized. To answer this question without reintroducing directly confined particles into the observable, we examine only the \textit{self} distribution of translationally free particles at equivalent internal stages of relaxation. This choice shifts the focus from an integrated measure to what the particles actually do.

\begin{figure*}[tb]
 \centering
 \includegraphics[width=0.91\textwidth]{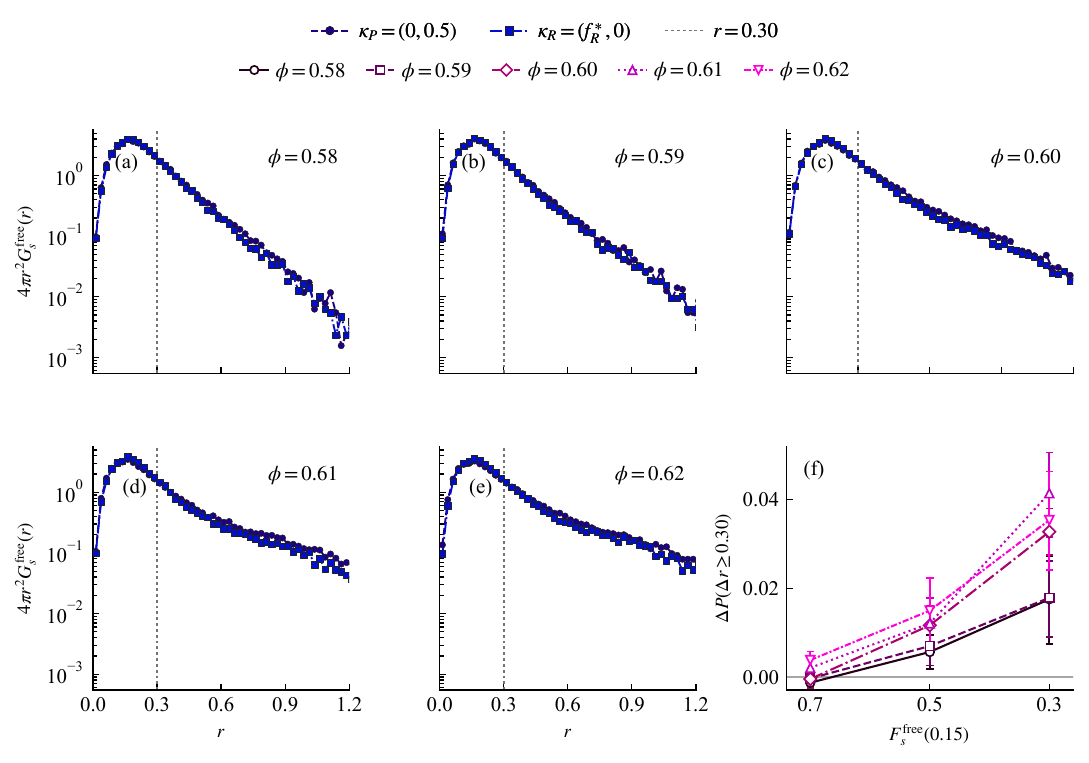}
\caption{\textbf{Redistribution of displacements among translationally free particles.} (a--e) Radial displacement distribution $p^{\rm free}(\Delta r) = 4\pi\Delta r^{2}G_s^{\rm free}(\Delta r)$ at the internal stage $F_s^{\rm free}(0.15)=0.30$ for $\ph=0.58$, $0.59$, $0.60$, $0.61$, and $0.62$. The vertical line marks the threshold $\Delta r=0.30$ used to define the tail. (f) Planar--pure-$\fR$ difference, $\Delta P_{\rm tail}$, in the probability that a free particle has $\Delta r\geq0.30$, evaluated at stages $F_s^{\rm free}(0.15)=0.70$, $0.50$, and $0.30$.}
 \label{fig:free}
\end{figure*}

Figs.~\ref{fig:free}(a)--\ref{fig:free}(e) show the radial displacement distribution $p^{\rm free}(\Delta r)$ at the stage $F_s^{\rm free}(0.15)=0.30$ for the five state points studied. Planar and pure-$\fR$ exhibit very similar distributions throughout the range, so the difference between the rules remains small at this microscopic level as well. To quantify specifically the region of larger displacements, we calculate the probability that a free particle has $\Delta r\geq0.30$ and define $\Delta P_{\rm tail}=P_P^{\rm free}-P_R^{\rm free}$. At this stage, $\Delta P_{\rm tail}>0$ at all five state points, showing that the planar dynamics yields a slightly higher probability of large displacements among translationally free particles. The magnitude of the effect remains small, consistent with the \emph{near-collapse} of the parametric relations, but it survives even after particles directly subjected to the planar constraint cease to contribute to the distribution.

Panel (f) shows that this excess of large displacements develops during relaxation. At $F_s^{\rm free}(0.15)=0.70$, $\Delta P_{\rm tail}$ is still small across the different state points. As the system progresses to stages $0.50$ and $0.30$, the difference increases and reaches its largest magnitude at the latest stage considered. The same pattern appears at all five values of $\ph$, although its amplitude varies among them. Thus, the difference between planar and pure-$\fR$ is not equally present throughout relaxation. It becomes progressively more apparent as large displacements account for a larger fraction of the dynamics.

This behavior connects the parametric residual to the characterization of dynamic heterogeneity through displacement distributions \cite{YamamotoOnuki1998,KobDonatiEtAl1997}. Recent results show that swap and local dynamics can have distinct van Hove statistics even when compared at similar global relaxation scales \cite{ShiraishiBerthier2024}. In the present case, the comparison is performed at common internal stages of relaxation, and only translationally free particles enter the distribution. Thus, the residual remaining after removal of the clock can be traced to a small but systematic difference in the probability of large displacements within this population.

\subsection{The forbidden direction leaves an imprint on the mobility of free particles}

\begin{figure*}[tb]
 \centering
 \includegraphics[width=0.95\textwidth]{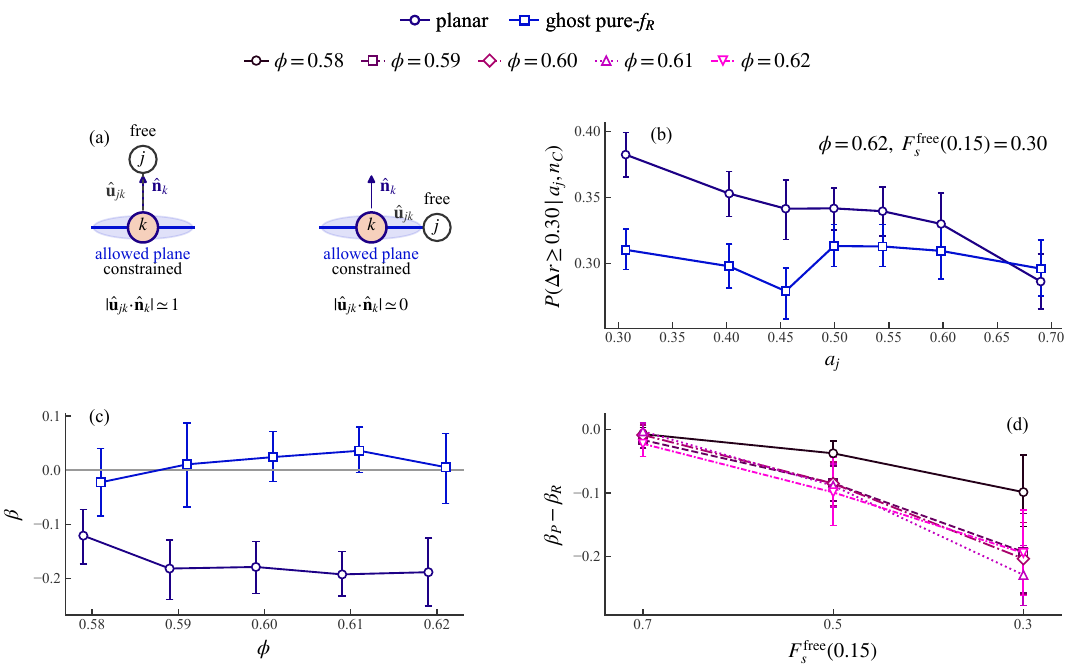}
 \caption{\textbf{Directional imprint of the planar constraint on translationally free particles.} (a) Geometry used to define orientational exposure. $\widehat{\bm n}_k$ is the forbidden normal of constrained particle $k$, and $\widehat{\bm u}_{jk}$ is the relative direction to free particle $j$. (b) Probability of $\Delta r\geq0.30$ as a function of $a_j$, with $n_C$ controlled, for $\ph=0.62$ at $F_s^{\rm free}(0.15)=0.30$. (c) Orientational slopes $\beta_P$ and $\beta_R$ for the planar dynamics and the pure-$\fR$ \emph{ghost} control, respectively, at the five state points and the same internal stage. (d) Difference $\Delta\beta=\beta_P-\beta_R$ across internal stages.}
 \label{fig:orientation}
\end{figure*}

The planar rule allows us to investigate an additional microscopic question because each constrained particle $k$ has a known direction, defined by the quenched normal $\widehat{\bm n}_k$, along which it cannot translate. For a free particle $j$, the relative direction $j$--$k$ can be nearly parallel or nearly perpendicular to this normal. In both cases, $j$ may have the same number $n_C$ of constrained neighbors; what changes is the orientation of its neighborhood relative to the directions along which those neighbors cannot move. Fig.~\ref{fig:orientation}(a) illustrates this geometry. We can therefore ask whether the mobility of free particles depends on this orientation even when the number of constrained neighbors is held fixed.

Fig.~\ref{fig:orientation}(b) shows that, in the planar dynamics, the probability that a free particle has $\Delta r\geq0.30$ decreases as the orientational descriptor $a_j$ increases, with $n_C$ controlled. In the pure-$\fR$ dynamics, the same identities and normals are retained only as \emph{ghost} labels, and the dependence of mobility on $a_j$ is approximately zero. Thus, the negative association between orientation and mobility observed under the planar rule persists after controlling for the number of constrained neighbors and is not reproduced when the same directions are used solely as geometric information, without restricting displacements.

This difference is evaluated using the slopes $\beta_P$ and $\beta_R$. As shown in Fig.~\ref{fig:orientation}(c), $\beta_P$ is negative at all five state points, whereas $\beta_R$ remains consistent with zero. The negative slope in the planar dynamics is also observed individually in 93 of the 100 configurations considered between $\ph=0.58$ and $0.62$. At $\ph=0.59$ and $0.62$, the three null tests (global-normal randomization, configuration-wise normal randomization, and permutation within fixed $n_C$) give distributions centered near zero, and none of the 200 realizations per test reaches the observed planar slope. These results associate the observed orientational dependence with the correspondence between local geometry and the directions actually forbidden by the planar rule.

The orientational dependence also develops during relaxation. Although the normals are quenched from the outset, the difference $\Delta\beta=\beta_P-\beta_R$ is small at $F_s^{\rm free}(0.15)=0.70$ and becomes progressively more negative by stage $0.30$ [Fig.~\ref{fig:orientation}(d)]. In other words, the geometry of the constraint is present from the initial configuration, but its association with the mobility of free particles becomes more pronounced as relaxation proceeds. Microscopic studies of supercooled liquids have previously identified directional organization of rearrangements and correlated motions \cite{KeysEtAl2011}. In the present case, the relevant direction is explicitly imposed by the kinetic rule, and the orientational dependence remains observable in particles without a translational constraint of their own.

\subsection{What remains after the clock is removed}

Returning to the analogy of Ghimenti \emph{et al.}, two dynamics corresponding to the same film projected at different speeds should exhibit the same relations between overlaps when the clock is removed \cite{GhimentiEtAl2026}. The near-collapse we observe shows that this description captures much of the correspondence between the planar and pure-$\fR$ rules. However, the curves do not coincide exactly. The small separation that remains indicates that reaching the same stage of relaxation at one spatial scale does not necessarily mean having reached the same stage at another. Therefore, changing the kinetic rule modifies, albeit modestly, the relation between the scales probed, and this difference cannot be absorbed into a simple change in the speed of evolution.

At comparable structural relaxation times, the planar and pure-$\fR$ rules differ in their departures from the common reference, as quantified by $\Delta\DTR$. The separation is also present in the direct, reference-free comparison through $\DPR$. Restricting the overlap to translationally free particles substantially reduces the difference but does not eliminate it, indicating that the residual extends beyond the direct contribution of particles confined to a plane. Consistently, the \textit{self} diagnostic reproduces most of the observed scale of the separation, whereas the \textit{distinct} diagnostic is considerably smaller.

The distribution of these displacements reveals how two rules with similar global relaxation can differ in the mobility of free particles. Under the planar dynamics, the probability of large displacements exhibits a small excess relative to pure-$\fR$, and the orientational analysis shows that this mobility also depends on neighborhood geometry. Even with the number of constrained neighbors held fixed, the probability of a large displacement decreases as the relative directions to neighboring particles become more aligned with their forbidden normals. This association is not reproduced by the \emph{ghost} control, in which the same normals are retained solely as geometric labels. Thus, a constraint applied directly to one population of particles leaves a measurable orientational dependence in the mobility of another population, whose translational displacements remain unrestricted. This observation adds a microscopic dimension to the distinction between the rules, although it does not, by itself, establish how much of the parametric separation can be attributed to the orientational dependence.

Our results thus help clarify the extent of \emph{time-reparametrization softness} in this system. Ghimenti \emph{et al.} showed that algorithmic changes can largely be absorbed by a transformation of the clock, with an empirical collapse that need not constitute an exact identity \cite{GhimentiEtAl2026}. Here, the family of rules introduced by Gavazzoni, Brito, and Wyart allows us to examine this correspondence while preserving the same equilibrium measure and modifying distinct kinetic channels \cite{GavazzoniBritoWyart2024}. We find a nearly preserved parametric relation, accompanied by resolved differences in the displacement distribution and by a dependence of free-particle mobility on the directions actually forbidden in its neighborhood. The orientational signature persists across the five state points investigated, between $\ph=0.58$ and $0.62$. Over this range, time reparametrization describes a broad correspondence between the dynamics, but the choice of kinetic rule remains apparent in how relaxation is distributed across scales and particles.

\FloatBarrier

\section{Conclusions}

Distinct kinetic rules can produce comparable structural relaxation
times and nearly coincident parametric relations between collective
overlaps, indicating that a change of clock captures much of their
common relaxation dynamics. A small but resolved separation
nevertheless remains after time is eliminated, persists when the
observables are restricted to translationally free particles, and
is also found in the direct comparison between planar and
pure-$\fR$. The self--distinct decomposition associates most of
this separation with the particles' own displacements, whose
distributions exhibit a modest redistribution toward larger
displacements as relaxation proceeds.

In the planar dynamics, even at fixed $n_C$, the probability that
a translationally free particle undergoes a large displacement
depends on the orientation of its neighborhood relative to the
quenched directions forbidden to constrained neighbors. This
association is absent within uncertainty in the ghost control and
is not reproduced by the null controls in the states tested.
Taken together, these results delineate the extent of
\emph{time-reparametrization softness} for the kinetic rules
investigated. A time-reparametrized description captures much of
their collective evolution, while measurable differences remain
in the displacement statistics of free particles and in the
dependence of their mobility on the local geometry of the imposed
constraint.

\renewcommand{\bibfont}{\footnotesize}
\bibliographystyle{apsrev4-2}
\makeatletter\immediate\write\@auxout{\string\citation{apsrev42Control}}\makeatother
\hypersetup{pdfborder={0 0 0}}
\let\originalhref\href
\renewcommand{\href}[2]{\originalhref{#1}{\textcolor{neonblue}{#2}}}
\renewcommand{\UrlFont}{\color{neonblue}\ttfamily}
\bibliography{artigo3_v5_bibcontrol,main}

\end{document}